\documentclass{rhumj_new}
\usepackage{listings}
\usepackage{color}
\usepackage{float}
\usepackage{siunitx}
\usepackage{graphicx} 
\definecolor{dkgreen}{rgb}{0,0.6,0}
\definecolor{gray}{rgb}{0.5,0.5,0.5}
\definecolor{mauve}{rgb}{0.58,0,0.82}
\title[The Case of Wisconsin]{Analyzing Swing States in Presidential Elections: The Case of Wisconsin}

\author[Zuo]{Michelle Zuo}
\affiliation{Thomas Jefferson High School for Science and Technology}
\email{michellepanda28@gmail.com}


\begin{document}

\abstract{This paper quantitatively analyzes county-level voting patterns in Wisconsin’s presidential elections from $2000$ to $2024$. As a pivotal swing state, Wisconsin has alternated between Democratic and Republican candidates since $2012$. Using data from the Wisconsin Elections Commission, we examine vote totals across $72$ counties and seven election cycles. Pearson correlations measure similarity in county voting trajectories, while choropleth maps visualize spatial shifts. Results show strong clustering of vote changes: Democratic and Republican gains between $2016$ and $2020$ were concentrated in southeastern urban and suburban counties, with rural areas showing little change. Correlations reveal a north-south divide, as southern counties exhibit similar trends and northern ones diverge. These findings highlight spatial heterogeneity in electoral dynamics and the decisive role of urban mobilization in statewide outcomes.}
\section{Introduction}

Presidential elections in the United States are ultimately decided not by the national popular vote, but by the Electoral College: a system in which each state's electoral votes are typically awarded on a winner-take-all basis. This structure transforms certain competitive states into crucial battlegrounds that can determine the outcome of the entire election. Wisconsin, with its history of narrow margins and recent electoral volatility, has emerged as one of the most critical swing states in modern American politics.

Understanding how voting patterns shift at the county level within swing states provides valuable insight into broader electoral dynamics. While state-level results capture headlines, county-level analysis reveals the geographic distribution of political change and can identify which communities are driving statewide trends. In Wisconsin, a state with diverse demographics ranging from urban Milwaukee to rural northern counties. These local patterns are particularly illuminating because small shifts in these pivotal areas can determine the entire state's electoral result.


Our analysis relates to the broader literature on swing state behavior and electoral competition. McLaren and Ma \cite{swing-state} developed a theoretical model demonstrating that electoral competition in systems like the U.S. Electoral College leads to policy outcomes that disproportionately favor swing states, those without strong partisan bias. Their "swing-state theorem" shows that under certain conditions, politicians will maximize the welfare of swing-state voters while largely ignoring non-swing-state preferences. They empirically validate this theory by examining U.S. tariff structures, finding that industries concentrated in swing states receive systematically higher protection. Their estimated parameter suggests the political process treats non-swing-state voters as worth approximately $70\%$ as much as swing-state voters.

While \cite{swing-state} focus on how swing-state status influences policy outcomes at the national level, our work examines the micro-level dynamics within a single swing state. Wisconsin's status as a critical battleground makes it an ideal case study for understanding how electoral competition manifests at the county level. By analyzing which Wisconsin counties have similar voting trajectories and how these patterns have shifted over recent election cycles, we provide complementary evidence about the geographic structure of political competition within swing states themselves.

Chen and Patel \cite{chen-patel} employed hierarchical clustering methods to analyze voting behavior across $15$ swing states in the $2020$ presidential election, connecting statistical patterns with political events such as the COVID-$19$ pandemic and the Black Lives Matter movement. They developed a Swing State Index to quantify electoral volatility and identified Wisconsin and Pennsylvania as a strongly connected pair in their clustering analysis, noting that healthcare concerns were a dominant factor in these states' voting behavior. My work extends this methodological approach by applying similar clustering techniques at a finer geographic resolution (the county level within Wisconsin) and across a longer time span ($2000-2024$ rather than just $2012-2020$). While\cite{chen-patel} examined state-level factors that influenced the $2020$ election outcome, we focus on understanding persistent voting patterns and similarities between counties over multiple election cycles, providing insights into the internal electoral geography of a key swing state.

The intersection of these research streams (the swing-state bias in electoral competition and the application of clustering methods to electoral data) provides a rich theoretical and methodological context for understanding Wisconsin's evolving electoral landscape. Elections are neither isolated events nor simple reflections of static preferences, but rather interconnected processes that reveal underlying patterns when examined over time and across geographic units.

This paper analyzes Wisconsin's presidential election results at the county level across seven election cycles from $2000$ to $2024$. Our primary objectives are threefold: first, to quantify how Democratic and Republican vote totals have changed over time in each of Wisconsin's $72$ counties; second, to identify counties with similar voting behavior patterns using correlation-based similarity metrics; and third, to visualize these trends through interactive choropleth maps that reveal geographic patterns in electoral shifts.

Our analysis relies on several key quantitative metrics. We calculate absolute and percentage changes in party vote totals between election cycles, with particular focus on the periods $2016-2020$, $2020-2024$, and the combined span from $2016-2024$. We compute win margins for each county and election year to track competitiveness and partisan lean over time. Additionally, we construct a county similarity matrix using Pearson correlation coefficients, allowing us to identify which counties exhibit similar voting trajectories despite potentially different geographic locations or demographic compositions.

The remainder of this paper is organized as follows. Section $2$ describes our data collection process, the structure of our datasets, and the derived metrics we calculated for analysis. Section $3$ explains our methodology for generating choropleth maps, including both vote change visualizations and county similarity maps. Section $4$ discusses potential future directions for this research, including analysis at the congressional district level and the application of Möbius inversion techniques to aggregate and disaggregate electoral data across different geographic boundaries. By combining rigorous quantitative analysis with intuitive geographic visualization, this study aims to provide a clearer picture of how Wisconsin's electoral landscape has evolved in recent decades.

\section{Election data}

\subsection{Data Collection}
Election data for this study were obtained from multiple authoritative sources to ensure comprehensive coverage of Wisconsin's presidential election results from $2000$ to $2024$. Current election data ($2020-2024$) was pulled directly from the official Wisconsin Elections Commission website\footnote{\url{https://elections.wi.gov/wisconsin-county-election-websites}} \cite {wisconsin_data} and individual county election offices. Historical election data for prior cycles ($2000-2016$) was compiled from Wikipedia's archived election results, which aggregate official state and county-level returns.

\subsection{Data Structure and Organization}

The raw election data were organized into a structured format with county-level vote totals for Democratic and Republican candidates across seven election cycles: $2000$, $2004$, $2008$, $2012$, $2016$, $2020$, and $2024$. For each county and election year, we recorded three primary variables: total votes cast, Democratic votes, and Republican votes.

To analyze electoral shifts over time, we calculated several derived metrics. For vote changes, we computed the difference in votes between consecutive election cycles:
\begin{equation}
\Delta V_{i,t}^{\text{party}} = V_{i,t}^{\text{party}} - V_{i,t-1}^{\text{party}}
\end{equation}
where $V_{i,t}^{\text{party}}$ is the vote total for a given party in county $i$ during election year $t$.

Percentage changes were calculated as the relative change in vote totals:
\begin{equation}
\text{Pct Change}_{i,t}^{\text{party}} = \frac{V_{i,t}^{\text{party}} - V_{i,t-1}^{\text{party}}}{V_{i,t-1}^{\text{party}}} \times 100\%
\end{equation}

Win margins were expressed as the difference between Democratic and Republican vote shares:
\begin{equation}
M_{i,t} = \frac{V_{i,t}^{\text{Dem}} - V_{i,t}^{\text{Rep}}}{V_{i,t}^{\text{Total}}} \times 100\%
\end{equation}
where positive values indicate Democratic victories and negative values indicate Republican victories.
\subsection{Derived Datasets}

From the primary election data, we generated four analytical datasets:

\subsubsection{Vote Change Dataset}
This dataset contains absolute and percentage changes in Democratic and Republican votes across multiple time periods. For each of Wisconsin's $72$ counties, the data includes total votes and party-specific votes for each election year, along with calculated differences for $2016-2020$, $2020-2024$, and $2016-2024$ periods. This structure enabled analysis of both short-term shifts between adjacent elections and longer-term trends spanning eight years.

\subsubsection{County Similarity Matrix}
To identify counties with similar voting patterns, we computed pairwise similarity scores between all counties using Pearson correlation coefficients. For any two counties $i$ and $j$, the correlation coefficient is defined as:
\begin{equation}
r_{i,j} = \frac{\sum_{t=1}^{n}(V_{i,t} - \bar{V}_i)(V_{j,t} - \bar{V}_j)}{\sqrt{\sum_{t=1}^{n}(V_{i,t} - \bar{V}_i)^2}\sqrt{\sum_{t=1}^{n}(V_{j,t} - \bar{V}_j)^2}}
\end{equation}
where $V_{i,t}$ represents a voting metric (such as Democratic vote share) for county $i$ in year $t$, and $\bar{V}_i$ is the mean across all years.

To make these correlation values more interpretable, we normalized them to similarity percentages:
\begin{equation}
S_{i,j} = \frac{r_{i,j} + 1}{2} \times 100\%
\end{equation}
which maps correlation values from $[-1, 1]$ to $[0\%, 100\%]$. The resulting matrix includes similarity percentages (normalized correlation values) and additional metrics such as Jaccard similarity for years in which counties flipped between parties:
\begin{equation}
J(A,B) = \frac{|A \cap B|}{|A \cup B|}
\end{equation}
where $A$ and $B$ are the sets of years in which the respective counties flipped between parties. 

\subsubsection{Win Margin Dataset}
For each county and election year, we calculated the winning party's margin of victory as both a percentage of total votes and an absolute vote difference. This dataset provides a clear view of competitiveness and partisan lean across Wisconsin's counties over time.

\subsubsection{Win Margin Heatmap Matrix}
To facilitate visual analysis of temporal voting patterns, we restructured the win margin data into a matrix format with counties as rows and election years as columns. Each cell contains the win margin percentage (positive values indicating Democratic wins, negative values indicating Republican wins), enabling rapid identification of county-level partisan shifts across multiple election cycles.

All datasets were cleaned and standardized to ensure consistency in county naming conventions and to handle missing or inconsistent values from source materials. We used Python scripts to verify data completeness across all $72$ Wisconsin counties for each of the seven election years analyzed. Since no counties have been added or removed from Wisconsin during the study period ($2000-2024$), we expected and confirmed complete coverage for all county-year combinations. We standardized county names by converting all entries to uppercase and removing trailing whitespace to ensure consistent matching with FIPS codes for geographic visualization. Any missing values (though none were identified in our dataset) would have been coded as N/A and displayed as gray regions in the choropleth maps to distinguish them from counties with zero or minimal vote changes.

\section{Generating maps} 
To visualize the electoral trends across Wisconsin counties, we created two different types of choropleth maps using Python's Plotly library. Choropleth maps are thematic maps in which geographic regions are colored or shaded according to a statistical variable—in this case, electoral margins and partisan lean. All maps were generated at the county level using Federal Information Processing Standard (FIPS) codes for geographic identification.

\subsection{Data Preparation}

For each visualization, we began by loading county-level election data from CSV files containing vote counts and calculated differences. To ensure accurate geographic mapping, we merged this data with Wisconsin County FIPS codes obtained from a public repository\footnote{\url{https://raw.githubusercontent.com/kjhealy/fips-codes/master/state_and_county_fips_master.csv}}. The merging process involved standardizing county names (converting to uppercase and removing trailing whitespace) and formatting FIPS codes as five-digit strings to maintain consistency with geographic boundaries.

\subsection{Vote Change Maps (2016-2020 and 2020-2024)}
We created separate choropleth maps tracking changes in Democratic and Republican vote totals across two election cycles. For each map, we calculated the absolute change in votes by party between elections, cleaning the data to handle comma-separated values and converting to a numeric format.

The visualizations employed custom color scales designed to reflect party affiliation while maintaining readability. Democratic vote changes used a blue gradient (ranging from dark navy to light cyan), while Republican changes used a red gradient (from deep maroon to light pink). We implemented binning endpoints using nine evenly spaced intervals between the minimum and maximum vote change values. The bin endpoints were calculated as:
\begin{equation}
b_k = V_{\min} + k \cdot \frac{V_{\max} - V_{\min}}{9}, \quad k = 0, 1, 2, \ldots, 9
\end{equation}
where $V_{\min}$ and $V_{\max}$ are the minimum and maximum vote changes across all counties. This approach allowed for a nuanced representation of the data distribution.

Using Plotly's \texttt{figure\_factory.create\_choropleth()} function, we generated county-level maps with white county outlines for clear geographic boundaries. Each map included a legend displaying the vote change ranges and was exported as an interactive HTML file for easy viewing and exploration.

\subsubsection{Republican Vote Change Analysis (2016-2020)}

\begin{figure}[h!]
\centering
\includegraphics[width=1\textwidth]{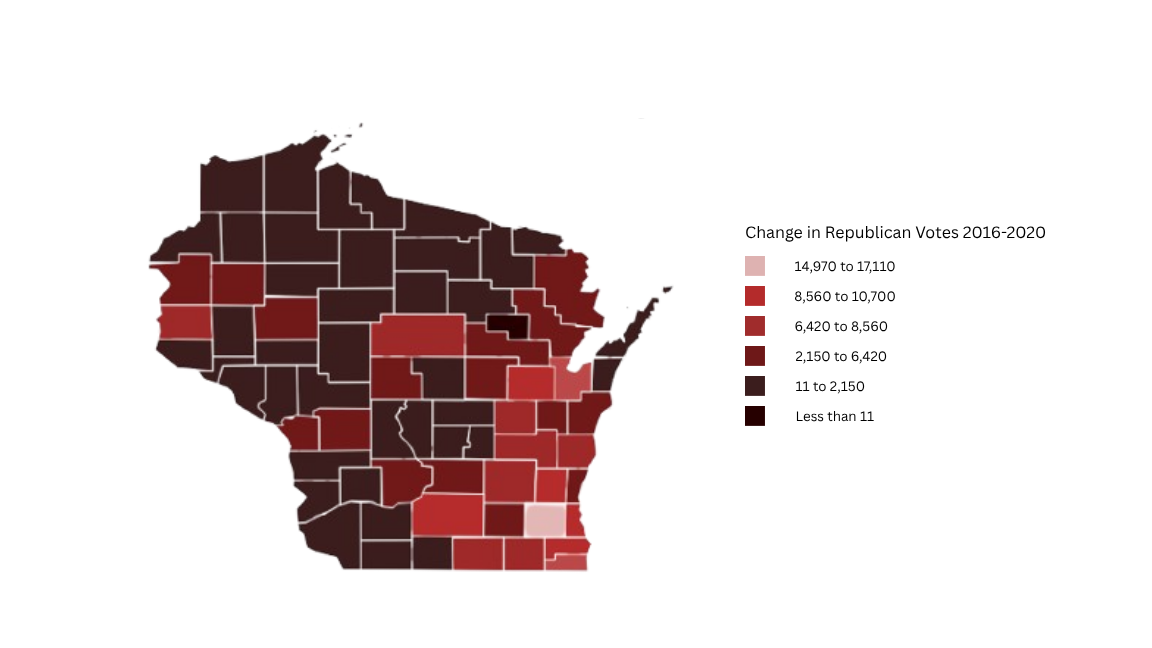}
\caption{Change in Republican Presidential Election Votes From 2016 to 2020}
\label{fig: Republican 2016-20 Vote Change Map}
\end{figure}

In Figure \ref{fig: Republican 2016-20 Vote Change Map}, Waukesha County stands out in the lightest shade of pink as the only county in the highest change category with an increase of $17,106$ votes. Counties bordering Waukesha and extending slightly northward also demonstrated significant growth in Republican votes, typically in the high thousands. This cluster includes major population centers such as Dane, Milwaukee, Kenosha, and Brown counties, suggesting increased Republican turnout in urbanized and suburban areas.

Two counties show notably high Republican vote increases despite being geographically separated from this southeastern cluster: St. Croix County in the west and Marathon County in central Wisconsin, both recording changes of approximately $5,000$. The majority of Wisconsin counties, particularly in the northern and western regions, experienced more modest increases in Republican votes, typically ranging from the low thousands to hundreds (represented by darker red shading). Menominee County stands out with the lowest vote change, indicating minimal growth in Republican turnout. This pattern suggests that Republican vote gains between 2016 and 2020 were concentrated primarily in Wisconsin's more populous counties and their surrounding areas, while rural counties saw relatively stable or minimal increases in Republican voter participation.

\subsubsection{Democratic Vote Change Analysis (2016-2020)}

\begin{figure}[h!]
\centering
\includegraphics[width=1\textwidth]{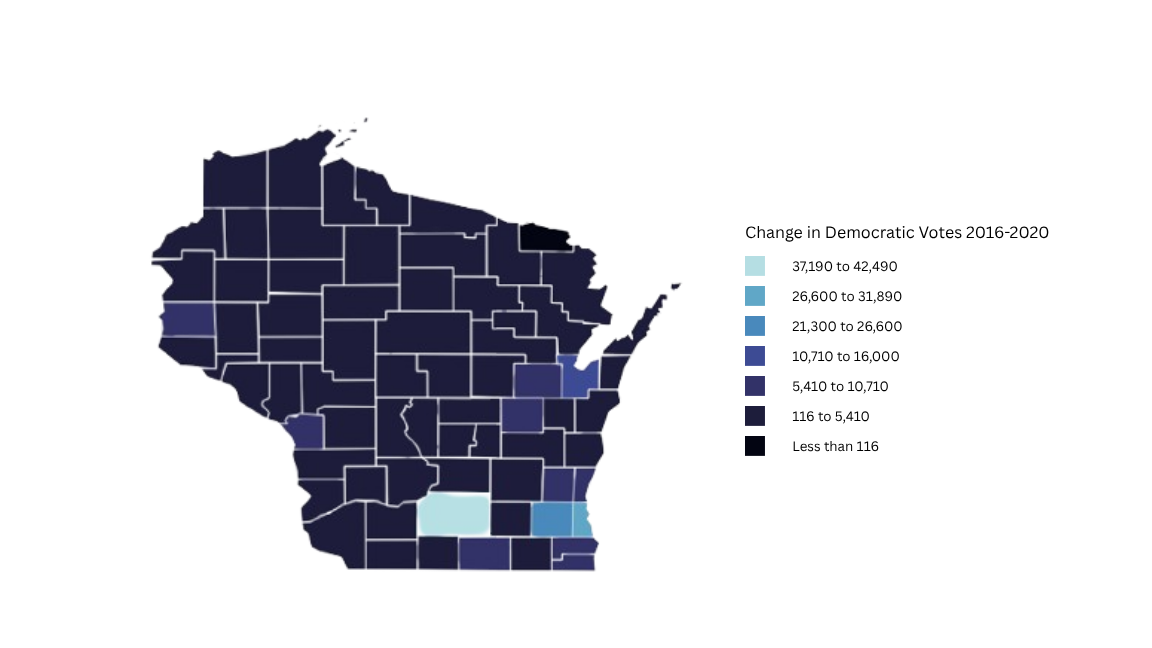}
\caption{Change in Democratic Presidential Election Votes From 2016 to 2020}
\label{fig: Democratic 2016-20 Vote Change Map}
\end{figure}
In Figure \ref{fig: Democratic 2016-20 Vote Change Map}, the concentration of Democratic gains in southeastern Wisconsin is striking, with Dane County appearing in the lightest blue as a clear outlier with approximately $42,500$ additional votes. Milwaukee and Waukesha counties follow as the second-tier growth leaders, each recording increases exceeding $20,000$ votes, forming a powerful southeastern urban corridor of Democratic mobilization.

The surrounding counties in southeastern Wisconsin, including Kenosha, Racine, and Ozaukee, also demonstrated significant Democratic vote growth in the mid-to-high thousands. Brown County in northeastern Wisconsin represents another notable concentration of increased Democratic turnout, suggesting urban mobilization extended beyond just the Milwaukee-Madison axis. In stark contrast, the vast majority of rural and northern counties experienced minimal Democratic vote increases, typically in the low hundreds to low thousands (represented by darker blue shading). Forest and Pepin counties recorded the smallest changes, with increases of approximately $100$ votes. This geographic distribution reveals that Democratic gains between $2016$ and $2020$ were overwhelmingly concentrated in Wisconsin's most populous urban and suburban counties, while rural areas saw comparatively little change in Democratic voter participation. This is a pattern even more pronounced than the Republican geographic concentration.

\subsection{County Similarity Map}
To examine voting pattern similarities across counties, we created a gradient choropleth map comparing each Wisconsin county to a reference county. Using precalculated pairs' similarity percentages from a separate analysis, we filtered the data to extract similarity scores relative to the target county, adding a self-similarity score of $100\%$ for the reference county itself.

This visualization employed Plotly Express's \texttt{px.choropleth()} function with the YlGnBu (Yellow-Green-Blue) color scale, which provided a smooth gradient representation of similarity values. Unlike the vote change maps, this approach used continuous color scaling rather than binned intervals, allowing for more granular visual differentiation between counties. The map was constrained to Wisconsin's geographic bounds using the \texttt{fitbounds} parameter for focused viewing.

\subsubsection{Dane County Similarity Analysis}

\begin{figure}[H]
\centering
\includegraphics[width=1\textwidth]{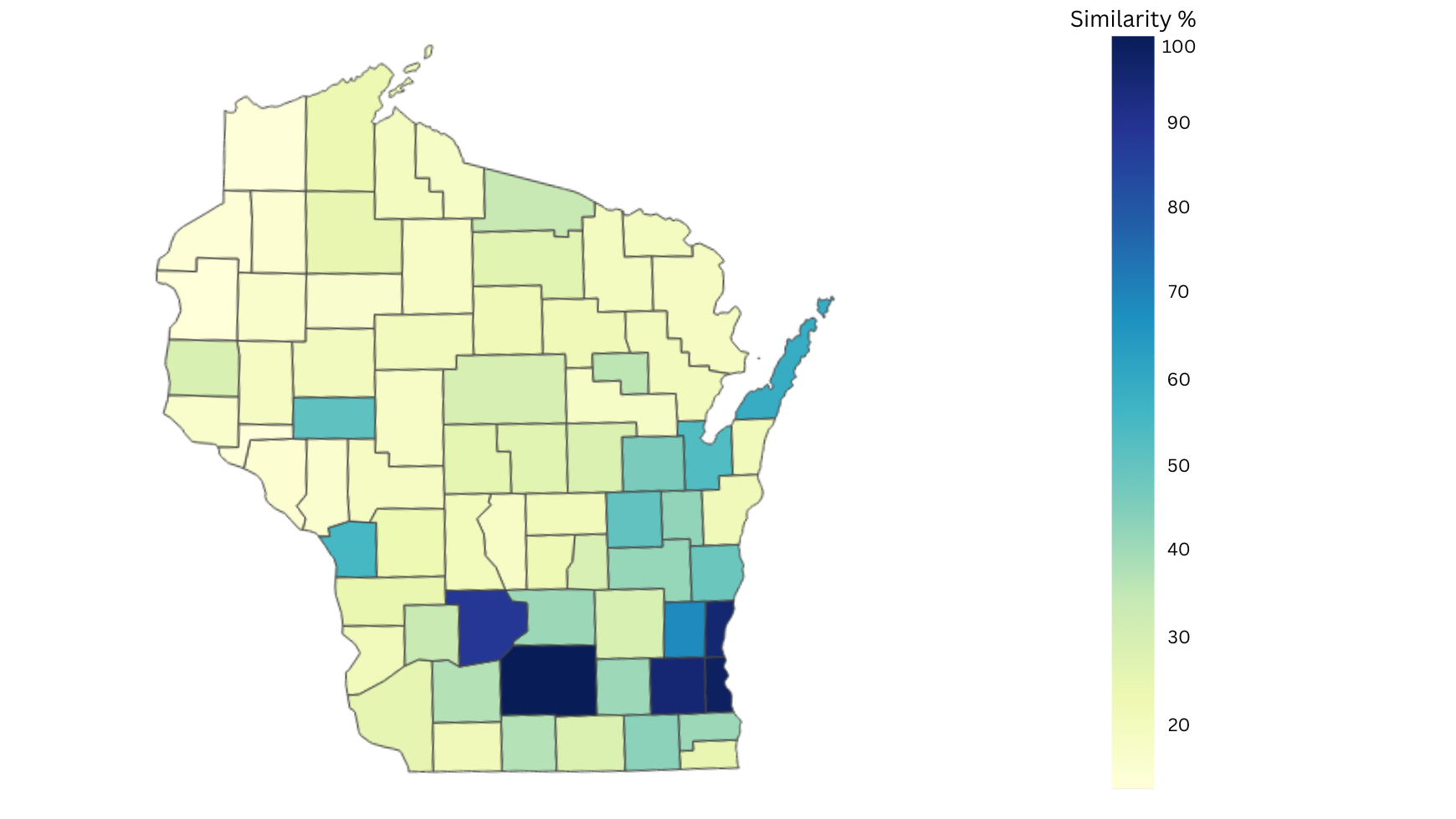}
\caption{Similarity of Wisconsin County's Election Pattern to Dane County}
\label{fig: Dane County Similarity Map}
\end{figure}

In Figure \ref{fig: Dane County Similarity Map}, the darker blue shading in southern Wisconsin contrasts sharply with the pale yellow coloring of northern counties, visually demonstrating the north-south political divide in the state. Counties exhibiting the highest similarity to Dane County (shown in darker blue, $80-100\%$ similarity) include Sauk, Milwaukee, Waukesha, and Ozaukee counties, suggesting that these areas share similar electoral trends with Dane County over the analyzed time period. In contrast, most counties in northern Wisconsin show significantly lower similarity scores (shown in pale yellow, $20-40\%$ similarity), indicating substantially different voting patterns. This geographic divide highlights a clear distinction between southern Wisconsin counties, which tend to follow similar electoral trajectories, and northern counties, which demonstrate more divergent political behavior.

All maps were saved as interactive HTML files, enabling users to hover over counties for detailed information and zoom in for closer inspection of specific regions.

\section{Summary and Future Work}

Although this study provides a comprehensive county-level analysis of Wisconsin's presidential elections, several promising directions remain for future investigation. These extensions would deepen our understanding of electoral dynamics and address some limitations of the current approach.

\subsection{Congressional District Analysis}

A natural extension of this work is to analyze voting patterns at the congressional district level. Unlike counties, which are fixed geographic units, congressional districts are explicitly designed for political representation and may better capture coherent communities of interest. Wisconsin's eight congressional districts each contain multiple counties, and analyzing how these districts have shifted over time could reveal patterns obscured by county-level aggregation. Furthermore, comparing presidential performance with congressional election results within the same districts could illuminate the extent of ticket splitting and party loyalty across different types of races.

\subsection{Hierarchical Aggregation via Möbius Inversion}

County-level data can contain significant noise, particularly in sparsely populated rural counties, where small absolute changes translate to large percentage swings. To address this, future work could employ the Möbius inversion function of combinatorics to systematically aggregate and disaggregate electoral data in different geographic hierarchies \cite{sagan-combinatorics}.

The Möbius function on a partially ordered set allows us to express relationships between data at different levels of aggregation. For a poset $P$ with partial order $\leq$, the Möbius function $\mu: P \times P \to \mathbb{Z}$ satisfies:
\begin{equation}
\sum_{x \leq z \leq y} \mu(x,z) = \delta_{x,y}
\end{equation}
where $\delta_{x,y}$ is the Kronecker delta.

In the context of electoral geography, we can define a poset where counties $\leq$ congressional districts $\leq$ state. If $f(x)$ represents the aggregate vote total over region $x$ and $g(x)$ represents the intrinsic contribution of region $x$ (excluding subregions), then Möbius inversion allows us to recover:
\begin{equation}
g(y) = \sum_{x \leq y} \mu(x,y) f(x)
\end{equation}

This technique could help isolate genuine regional trends from noise and enable meaningful comparisons between cities, counties, and congressional districts at different scales of analysis.

\subsection{Optimal Transport for Vote Flow Analysis}

An innovative approach to understanding electoral shifts would be to apply optimal transport theory to model how votes "move" between parties and across geographic regions between elections \cite{peyre-cuturi}. While individual voters' choices are private, optimal transport provides a mathematical framework for inferring the most likely aggregate patterns of vote transfer.

For two probability distributions $\mu$ and $\nu$ (representing vote distributions in consecutive elections), the optimal transport distance (Wasserstein distance) measures the minimal "cost" of transforming one distribution into another:
\begin{equation}
W_p(\mu, \nu) = \left( \inf_{\gamma \in \Pi(\mu,\nu)} \int_{X \times Y} d(x,y)^p \, d\gamma(x,y) \right)^{1/p}
\end{equation}
where $\Pi(\mu,\nu)$ is the set of all transport plans with marginals $\mu$ and $\nu$, and $d(x,y)$ is a ground distance (e.g., geographic distance between counties).

By computing optimal transport plans between election cycles, we could identify which counties gained or lost voters to which parties, and whether these changes correlate with geographic proximity, demographic shifts, or other factors. This approach would provide a more nuanced view of electoral dynamics than simple vote difference calculations. Recent work by Lanzetti et al. (2022) demonstrates how Wasserstein gradient flows can model the evolution of political party ideologies over time, providing a theoretical foundation for applying optimal transport methods to electoral analysis \cite{lanzetti-political}.

\subsection{Temporal Modeling and Forecasting}

Future work could also incorporate time series analysis to model temporal trends in county-level voting patterns. Techniques such as autoregressive models, state-space models, or even machine learning approaches could be applied to forecast future election outcomes based on historical trends. Understanding the autocorrelation structure of county-level vote shares over time would reveal whether electoral shifts are persistent or mean-reverting.

\subsection{Demographic and Socioeconomic Integration}

Finally, integrating demographic and socioeconomic data with electoral results would provide explanatory power beyond pure vote tallies. Variables such as population growth, median income, educational attainment, racial composition, and age distribution could be incorporated into regression models to identify which factors best predict electoral shifts. This would transform the analysis from purely descriptive to explanatory, offering insights into the underlying mechanisms driving Wisconsin's evolving electoral landscape.

\subsection{Conclusion}
This paper has presented a comprehensive county-level analysis of Wisconsin's presidential elections from 2000 to 2024. We use correlation-based similarity metrics and choropleth visualizations to reveal geographic patterns in electoral behavior. Our analysis demonstrates that Wisconsin counties cluster into distinct groups with similar voting trajectories, with a clear divide between southern counties that share similar electoral trends and northern counties that exhibit more divergent patterns. The proposed extensions, ranging from congressional district analysis and Möbius inversion techniques to optimal transport methods and demographic integration, would substantially broaden the scope of this work, moving from descriptive statistics toward predictive modeling and causal inference about the mechanisms driving electoral change in this critical swing state.

\section{Acknowledgment}
I would like to thank the Institute for Mathematics and Democracy (IMD) for providing the opportunity to participate in their high school summer research program, where this work was initiated. I am especially grateful to Professor Natasa Dragovic for her invaluable guidance and support throughout the development of this paper. I would also like to acknowledge my fellow high school participants in the IMD program for their collaboration during the summer research experience.

\end{document}